\documentclass[aps,floats,twocolumn,superscriptaddress,prb]{revtex4}
\usepackage{amsmath}

\usepackage{epsfig}
\usepackage{graphicx}
\usepackage{caption}
\usepackage{dcolumn}
\usepackage{bm}

\newcommand{\beq}{\begin{equation}}
\newcommand{\eeq}{\end{equation}}
\newcommand{\beqa}{\begin{eqnarray}}
\newcommand{\eeqa}{\end{eqnarray}}
\def\gapp{\lower.35em\hbox{$\stackrel{\textstyle>}{\sim}$}}
\def\lapp{\lower.35em\hbox{$\stackrel{\textstyle<}{\sim}$}}

\begin{document}
\bibliographystyle{naturemag}
%


\title{Hall viscosity from elastic gauge fields in Dirac crystals}

\author{Alberto Cortijo}
\affiliation{Instituto de Ciencia de Materiales de Madrid,\\
CSIC, Cantoblanco; 28049 Madrid, Spain.}

\author{ Yago Ferreir{\'o}s}
\affiliation{Instituto de Ciencia de Materiales de Madrid,\\
CSIC, Cantoblanco; 28049 Madrid, Spain.}
\author{ Karl Landsteiner}
\affiliation{Instituto de F\'isica Te\'orica UAM/CSIC, \\
Nicol\'as Cabrera 13-15, Cantoblanco, 28049 Madrid, Spain }

\author{Mar\'{\i}a A. H. Vozmediano}
\affiliation{Instituto de Ciencia de Materiales de Madrid,\\
CSIC, Cantoblanco; 28049 Madrid, Spain.}

\date{\today}
\begin{abstract}
The combination of Dirac physics and elasticity has been explored at length in graphene where the so--called "elastic gauge fields" have given rise to an entire new field of research and applications: Straintronics. The fact that these elastic fields couple to fermions as the electromagnetic field, implies that many electromagnetic responses will have elastic counterparts not explored before. In this work we will first show that the presence of elastic gauge fields will be the rule rather than the exception in most of the topologically non--trivial materials in two and three dimensions. In particular we will extract the elastic gauge fields associated to the recently observed Weyl semimetals, the "three dimensional graphene". As it is known, quantum electrodynamics suffers from the chiral anomaly whose consequences have been recently explored in matter systems. We will show that, associated to the physics of the anomalies, and as a counterpart of the Hall conductivity, elastic materials will have a  Hall viscosity in two and three dimensions with a coefficient  orders of magnitude bigger than the previously studied response. The magnitude and generality of the new effect will greatly improve the chances for the experimental observation of this topological, non dissipative response.
\end{abstract}
%
%
%
%
 \maketitle

\section{ Introduction}
In elasticity theory, Hooke's law tells us that the stress tensor $T_{ij}$ applied on an elastic solid is proportional to the strain tensor $u_{ij}=(\partial_{i}u_{j}+\partial_{j}u_{i})/2$ as $T_{ij}=\lambda_{ijlr}u_{lr}$ through a four-rank tensor $\lambda_{ijlr}$. If the system is viscoelastic the stress tensor is also proportional to the \emph{time derivative} of $u_{ij}$:
\beq
T_{ij}=\eta_{ijlr}\dot{u}_{lr},
\label{HV}
\eeq 
through another four-rank tensor $\eta_{ijlr}$. In general, this viscosity tensor possesses both symmetric and antisymmetric components under the permutation of pairs of indices. While the symmetric part is generally associated to dissipation and vanishes at zero temperature, the antisymmetric part arises when time reversal symmetry is broken\cite{ASZ95}. As usually happens with transport coefficients like the gyrotropic term of the permittivity in dielectrics, the antisymmetric part of the coefficients that are odd under time inversion are dissipation-less and they are not constrained to vanish at zero temperatures. In two dimensions, isotropy, the symmetry under permutations of $ij$ and $lr$ indices, and the aforementioned antisymmetry in permutations of pairs of indices allow for only one independent element of the antisymmetric part of the tensor $\eta^{H}_{ijlr}$: $\eta_{ijlr}=\frac{\eta^H}{2}(\delta_{il}\epsilon_{jr}+\delta_{ir}\epsilon_{jl}+\delta_{jl}\epsilon_{ir}+\delta_{jr}\epsilon_{il})$, where $\epsilon_{ij}$ is the Levi-Civita tensor in two dimensions.

The first context where this antisymmetric, dissipation-less coefficient was described was the quantum Hall effect (this is why this coefficient is called Hall viscosity)\citep{ASZ95}. Being non-dissipative and appearing in quantum Hall effect systems, it is not strange that the Hall viscosity is associated to a certain Berry phase endowed with a topological meaning. 
In the case of the Hall conductivity, the off diagonal part of the conductivity tensor $\sigma_{xy}$ is related to a non-zero Berry phase associated to the evolution of the phase of the wave function in the Brillouin zone\cite{TKKN82}. In the case of the Hall viscosity, the parameter space is the space of the displacements of the atoms in the continuum limit, $\bm{u}(\bm{r})$. Since these displacements are considered adiabatic with respect to the electronic motion, one can consider a cyclic evolution of the wave function in this particular parameter space and define an associated Berry curvature of the wavefunction $\Phi_{a}$ at some energy level $\varepsilon_{a}$: $\Omega^{(a)}_{ijlr}=i\langle\partial_{u_{ij}}\Phi_{a}|\partial_{u_{lr}}\Phi_{a} \rangle$. 

Also the Hall viscosity can be computed from a Kubo formula\cite{BGR12} in terms of the stress-stress correlation function $\eta_{ijlr}=\int dt e^{i 0^{+}t}\langle[T_{ij}(t),T_{lr}(0)]\rangle$ with the stress tensor defined as the variation of the Hamiltonian with the strain tensor $T_{ij}=\frac{\delta H}{\delta u_{ij}}$. After some mathematical manipulations, a part of this Kubo formula appears to be proportional to the Berry curvature $\Omega_{ijlr}$ defined above.

Besides, as it happens in all transport coefficients, the Hall viscosity appears in the effective action of some low energy degrees of freedom after integrating out the fermionic degrees of freedom. The effective energy functional for elastic displacements (phonons after quantizing) for a system in two spatial dimensions consists on two terms, the standard elastic terms (possibly renormalized by the effect of electrons) and the Hall viscosity term:
\begin{equation}
\mathcal{U}=\frac{1}{2}\int d^{2}\bm{r} [\lambda_{ijlr}\partial_{i}u_{j}\partial_{l}u_{r}+\eta^{H}_{ijlr}\partial_{i}u_{j}\partial_{l}\dot{u}_{r}].
\end{equation}

Today there is a consensus about the meaning and the origin of the Hall viscosity. The question at this point is how to define the coupling between electrons and the elastic displacements $u_{i}$ or better, to the strain tensor $u_{ij}$ and where to seek for it. The quantum Hall state is a topologically ordered state so it is natural to extend the search to other topologically non-trivial quantum systems. Much work has been done in this direction and Hall viscosities have been found in  most of the topologically nontrivial condensed matter phases apart from the quantum Hall phase\citep{ASZ95} like the fractional quantum Hall phase\cite{Read09,GGA15}, chiral superfluids\cite{Read09,HMS14} and superconductors\cite{SK14}, Chern insulators in presence of torsion in two dimensions\cite{HLF11,HLP12,SW14} and, more recently, in three dimensions\cite{SHR15,SW14} as well. 

Precisely, the step of finding topologically non trivial systems from two to three spatial dimensions appears to be rather challenging. To have a non-zero Hall viscosity in three spatial dimensions  breaking time reversal symmetry is not enough. Spatial isotropy or rotational invariance in the continuum limit must be broken as well\citep{ASZ95}. Up to now, only the torsional Hall viscosity defined in Weyl semimetals\cite{SW14} and three dimensional chiral superconductors\cite{SK14} has been discussed in the literature. In these systems torsion is linked to the presence of dislocations\cite{HLF11} that define a preferred direction in the space. Also, in three dimensional topological insulators a Hall viscosity is allowed  when the spatial isotropy is broken by the presence of sample surfaces, and time reversal symmetry is broken by coupling magnetic elements to the two dimensional states appearing at these surfaces\cite{SHR15}.

As we will discuss in detail later, the way how the elasticity couples to the low energy degrees of freedom (electrons or phase fluctuations in the case of superfluids or superconductors) is crucial to determine if a topologically nontrivial phase displays Hall viscosity. Most of the studies available in the literature are based on a continuum formulation of the system and the elastic degrees of freedom. The philosophy is that, since elastic deformations can be viewed as geometrical deformations in the medium hosting the system, the excitations feel a distorted or curved space where to propagate. We can thus define an effective metric tensor related to elastic distortions and postulate that our system now develops its dynamics in a curved space. The general formalism  of field theories in curved spaces and linear responses does the job\cite{Hoyos14}. Although powerful, this approach has some limitations in the scope of systems that can be treated. For instance, in the metric formalism (in absence of torsion), some non trivial topological phases with non-zero Hall conductivity have zero Hall viscosity. This is the case of 
Chern insulators in two dimensions and Weyl semimetals in three dimensions.
On the other hand, only a small fraction of the literature devoted to the Hall viscosity in electronic systems treats the problem from the more conventional perspective in solid state physics through general electron-phonon couplings\cite{BCQ12,SHR15}. The goal in these approaches is to define an effective electron-metric coupling by taking the continuum limit from a lattice formulation, thus reaching similar success and suffering from the same limitations as in the former case.
\\

The present work offers a different alternative. We will show how the standard electron-phonon coupling in a class of electronic systems with an underlying  lattice not only gives rise to a metric formulation in the continuum, but as we know from graphene, emergent vector fields arise in such effective description coupling to the electrons in a very similar way as the electromagnetic gauge field. These vector couplings between electrons and elastic deformations offer new results and enlarge the kind of systems where the Hall viscosity can in principle be measured.


\section{ Hall viscosity from Hall conductivity in graphene}


Dirac materials is the generic name for electronic systems whose low energy excitations are described by a Dirac Hamiltonian, graphene being the paradigmatic example. They exist in two and three spacial dimensions and have been the focus of attention in last years condensed matter. 
We will first show that anomalous Hall conductivity implies anomalous Hall viscosity in graphene and we will then extend the discussion to other Dirac materials.

The low energy action around one of the Fermi points of graphene coupled to a U(1) electromagnetic background field  is 
\beq
S=\int d^3 x \bar\psi \left(i\gamma^{0}\partial_{0}-iv_{F}\gamma^i \partial_i+\frac{ev_{F}}{c} \gamma^{i}A_i\right)\psi.
\eeq
A way to break time reversal invariance suggested by Haldane is to add complex next-to-nearest neighbours in the tight-binding on the honeycomb lattice\cite{H88}. In the continuum limit, these complex hoppings lead to a mass term with opposite sign at each Fermi point. After integrating out fermions it is easy to find the Chern Simons term as a part of the effective electromagnetic action:
\beq
S_{CS}=\frac{e^{2}}{\pi c}sign(m) \int d^3 x\epsilon^{ijk} A_i\partial_j A_k,
\label{CSaction}
\eeq
after considering all the degeneracies (and in units of $h=1$). The term (\ref{CSaction}) is a consequence of the parity anomaly. Coupling the effective action \eqref{CSaction} to a background electromagnetic source $J^i A_i$ allows to derive  an anomalous quantum Hall effect:
\beq
J^i\equiv\frac{\delta S}{\delta A_i}=\frac{2e^2}{\pi c}sign(m)\epsilon^{ijk}\partial_j A_k.
\eeq

A very interesting aspect of graphene shared by many other 2D materials is the tight connection between lattice structure and electronic excitations. Within the simplest tight binding-elasticity approach, lattice deformations couple to the Dirac fermion as elastic U(1) vector fields that depend linearly on the strain tensor $u_{ij}= \partial_i u_j +\partial_j u_i$ as \cite{KM97,SA02b}
(see section Methods \ref{elgraphene})
\begin{eqnarray}
A^{el}_1 &=&\frac{\beta }a\left( u_{11}-u_{22}\right) ,  \nonumber
\\
A^{el}_2 &=&-\frac{2\beta }au_{12}, \label{strainfield}
\end{eqnarray}
where $\beta$  is the dimensionless Gr\"{u}neisen parameter estimated to be of order 2 in graphene, and $a$ is the lattice constant.  Since lattice deformations do not break time reversal symmetry, the elastic vector field couples with opposite signs to the two Dirac points and behaves as an axial vector field in a four dimensional representation. 

If we consider this elastic axial vector coupling to electrons instead of the electromagnetic field we have
\beq
S_g=\int d^3 x \bar\psi \left(i\gamma^{0}\partial_{0}-iv_{F}\gamma^i \partial_i+v_{F} \gamma^{i}A^{el}_i\right)\psi+m\bar\psi\psi,\label{firsterm}
\eeq
It is obvious that the same line of arguments as before will give rise to a Chern-Simons action like \eqref{CSaction} in terms of the elastic instead of the electromagnetic field. In particular we will get 
\beq
S_{CS}=\frac{2}{\pi}sign(m)\int d^3 x \left(A^{el}_1\dot {A}^{el}_2-A^{el}_2\dot{A}^{el}_1\right),
\label{CS2D}
\eeq
substituting from eq. \eqref{strainfield} we get in the Chern Simons action a term of the form
\begin{eqnarray}
S_{CS}[u]&=&\frac{4\beta^2}{\pi a^2}sign(m) \int d^3 x [u_{22}\dot{u}_{12}-u_{11}\dot{u}_{12}+\nonumber\\
&+&u_{12}\dot{u}_{11}-u_{12}\dot{u}_{22}].
\end{eqnarray}
To this effective action we can now couple a source term for a time-dependent elastic deformation $S_u=u_{ij}T^{ij}$ and compute the averaged value of the $11$ component of the stress tensor
\beq
\langle T_{11}\rangle\equiv \frac{\delta S}{\delta u_{11}}
=\frac{4\beta^2}{\pi a^2}sign(m)\dot{u}_{12}. 
\eeq
Hence we get 
the quantized Hall viscosity  
\beq
\eta^H=\frac{4\beta^2}{\pi a^2}sign(m).
\label{hv}
\eeq
The coefficient is determined by the parity anomaly in (2+1) dimensions, and the characteristic length associated to the viscosity is given by the lattice spacing  $a$. This new response is a genuine viscoelastic response: Elastic gauge fields inducing a Hall viscosity.

The Hall viscosity described here is an intrinsic property of the Haldane model of graphene in the same sense as the quantum anomalous Hall effect (QAHE) of the original proposal \cite{H88}.  It has been argued \cite{Kimura00} that the  intrinsic Hall viscosity in the topological phase of the Haldane model is $\eta_H=\hbar\rho/4$, where $\rho$ is the electronic density. When the Fermi level lies within the gap, no Hall viscosity appears in the Haldane model.

Although the Haldane model has been realised experimentally in optical lattices \cite{JMetal14} 
it seems challenging to generate it in graphene. The Haldane model is the model of the QAHE in graphene and the QAHE by magnetic proximity effect has been measured \cite{QRetal14}, making graphene a suitable platform to potentially measure effects induced by the presence of the Hall viscosity.

The same reasoning as done in the Haldane model can be applied to real graphene in a perpendicular magnetic field breaking time reversal symmetry. The  Hall conductivity in graphene is \cite{Jackiw84,GS05} $\sigma_H= 4e^2 (n+1/2)/\pi, n=0, 1, ..$. By the same argument as before in the quantum limit where only the lowest Landau level is filled ($n=0$), there is a contribution to the Hall viscosity coming from the elastic vector fields:
\beq
\eta_H\sim\frac{2\beta^{2}}{a^{2}}.
\label{HVHel}
\eeq

The complete low energy action for deformed graphene in the elastic limit has been derived in \cite{MJSV13}. In the  presence of a magnetic field other terms coupling the elasticity to the Dirac fermions contribute to the Hall viscosity.
%
In particular the following term 
\beq
H=\frac{i}{2}\beta u_{ij}\left(\psi^{+}\sigma_{i}\partial_{j}\psi+\partial_{j}\psi^{+}\sigma_{i}\psi\right),\label{secondterm}
\eeq
gives a contribution to the Hall viscosity that has been computed to be \cite{ASZ95,Kimura00}
\beq
\eta^{(B)}_H=4\beta^{2} [\vert n\vert (\vert n\vert +1)/4+1/8] e B/2\pi.
\label{HVH}
\eeq
In the quantum limit ($n=0$) we thus have $\eta_H^{(B)}\sim\beta^2/ l_B^2$,  where $l_B$ is the magnetic length. This result might seem a little bit puzzling at a first glance. We have stressed that in the absence of torsional couplings and magnetic fields, the Haldane model has no Hall viscosity unless one considers the elastic gauge fields. The reason is that the coupling in (\ref{secondterm}) is a derivative coupling. When one consider both terms (\ref{firsterm}) and (\ref{secondterm}) and find the effective theory for elasticity, the term coming from (\ref{secondterm}) has more than one space-time derivatives and it does not contribute to the Hall viscosity. When a magnetic field is taken into account, the partial derivative $\partial_{j}$ transforms into a covariant derivative $\partial_{j}+ieA_{j}$ (with $\vec{B}=\nabla\times\vec{A}$) and the magnetic length $l_{B}=\sqrt{c/eB}$ permeates through the calculations allowing for the result (\ref{HVH}).
Taking graphene as an example, the ratio between the two contributions is
\beq
\frac{\eta_H}{\eta_H^{B}}\sim\frac{l^{2}_{B}}{a^{2}}\sim\frac{10^4}{B},
\eeq
where $a$ is the lattice constant and  $B$ is the magnitude of the magnetic field in Tesla. For a standard magnetic field of $10$T, the contribution from the elastic gauge fields is three orders of magnitude larger than the one coming from (\ref{secondterm}). It means that if the Hall viscosity is measured in graphene under quantizing magnetic fields, it is the component discussed in the present work the one that will be measured. 

\section{Generality of the effect }
\begin{figure}
\includegraphics[scale=0.3]{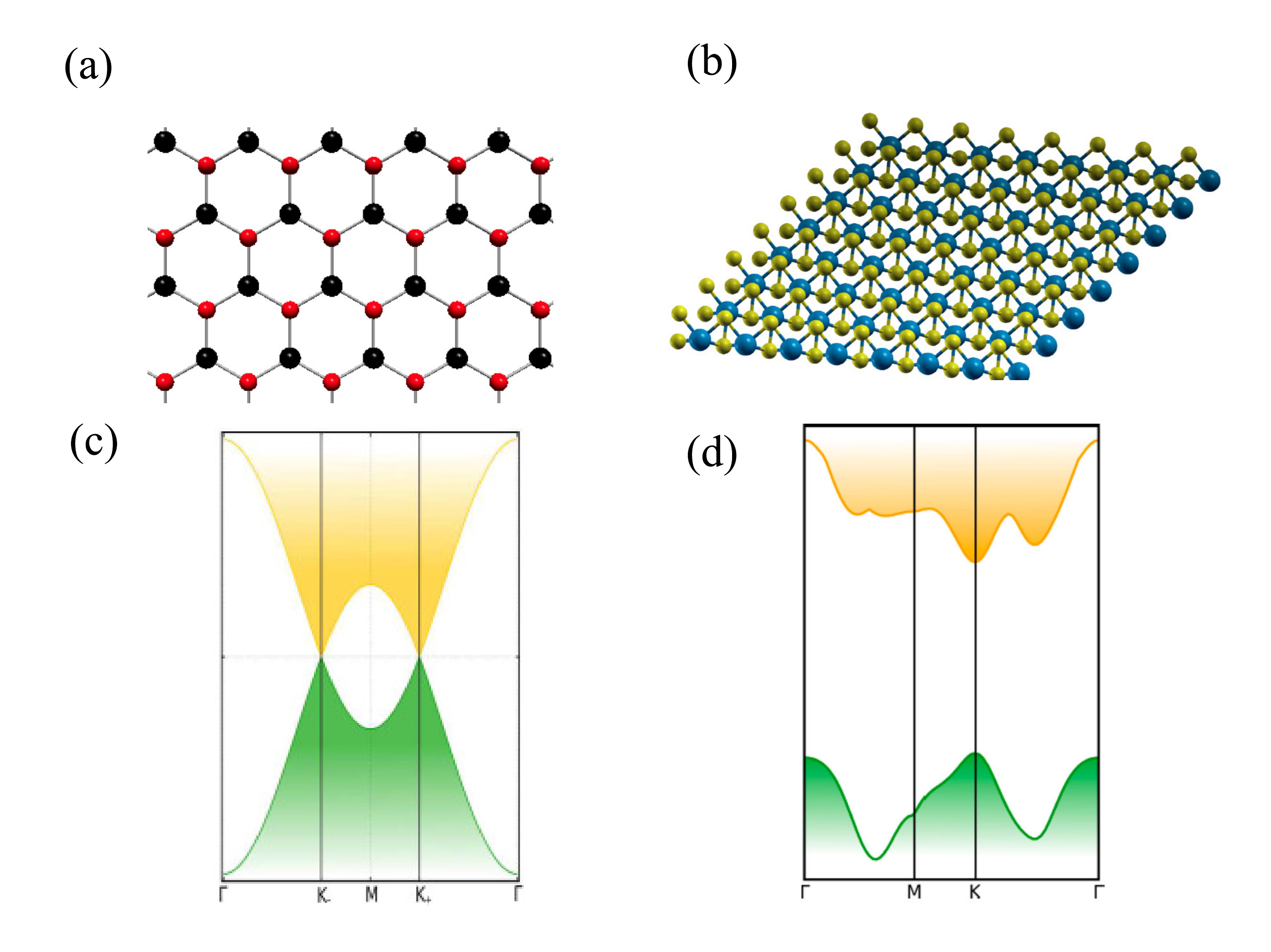}
\caption{\textbf{Two dimensional Dirac materials}. The most representative example of the so-called Dirac materials in two spatial dimensions is graphene. In graphene, carbon atoms are arranged in a honeycomb lattice, (a). The highest occupied and the lowest empty bands touch at the corners of the Brillouin zone and the low energy band dispersion is well described by two species of massless Dirac fermions (c). MoS$_{2}$ is the representative of the family of transition metal dichalcogenides, (b). MoS$_{2}$ is a semiconductor with bands that admit a low energy effective description in terms of two species of massive Dirac fermions (d).}
\label{2Dbands}
\end{figure}
The novel effect described in graphene is based on two main ingredients: The existence of elastic gauge fields, and the non--trivial topology  of the electronic bands of the system. This last ingredient is what allows to write down a Chern Simons effective action in 2D given in eq. \eqref{CSaction} which is the key issue of the approach (D will denote the spacial dimensions of the (D+1) system). The non trivial topology can arise directly as a Berry phase associated to the Bloch bands of the crystal as in the Haldane model \cite{H88}, or from the topology of a general 2D electron gas in a perpendicular magnetic field through the Hofstadter model \cite{Hofs76,TKKN82}. 
In all cases the Hall conductivity is associated to the topological Chern--Simons action depicted in eq. \eqref{CSaction}. As we have demonstrated, a Hall conductivity and elastic gauge fields inmediately implies the new Hall viscosity. We will now discuss the generality of the presence of elastic gauge fields in real systems and give some concrete examples.

Although first derived in a tight binding formulation on the lattice, the generation of pseudogauge fields coupling shape deformations to electronic degrees of freedom in the low energy effective Hamiltonian of electron systems, is a very general phenomenon. It also arises in a  general construction of low energy effective actions based on a symmetry analysis as described in \cite{Ma07,JSV12,JMV13}. The general conditions for a 2D lattice to support elastic gauge fields have been discussed in \cite{Ma07}. The two essential ingredients are: The presence of (at least) two atoms per unit cell  (this can arise from geometry as in graphene, from orbital degrees of freedom, or by other mechanisms). Lack of inversion symmetry in the little group leaving a Fermi point invariant, ensures that the Fermi points sit at non-equivalent high symmetry points of the Brillouin zone.  It is easy to see that the same structure giving rise to the Dirac Hamiltonian causes the minimal coupling to the vector fields associated to lattice deformations implying that the phenomenon can be generalized to other crystals in two and three dimensions.  As it is clear from the analyses in \cite{Ma07,JMV13} the symmetry group of the lattice dictates the precise form of the elastic vector fields.


In 3D, an analysis of all crystallographic groups that sustain Weyl points in its low energy action in complete analogy  to the case of graphene was first presented in ref. \cite{Manes12}. Since the Dirac structure in these crystals comes from the lattice, these having Weyl points  in non--equivalent high symmetry  points of the Brillouin zone will probably generate  elastic vector field couplings under strain although a general analysis as the one done in 2D is still lacking. 

\subsection{2D and  Van der Waals systems}

New two dimensional materials of scientific and technological interest are being synthesized in the ``post--graphene" era \cite{RCetal15}. They share many of the properties of graphene and often have large gaps making them more suitable for electronic applications.  Single-layer and multilayer transition metal dichalcogenides $MX_2$ ($M=$ Mo, W and $X=$ S, Se) \cite{RCetal15} do support elastic gauge fields. Although having a noticeable band gap and a complicated orbital structure, $MoS_2$ and the related compounds based on the honeycomb lattice, support at low energy, the same elastic gauge fields as  graphene given in eq. \eqref{strainfield} \cite{COG14}. These are also found in bilayer graphene \cite{MPO12} which has a quadratic dispersion  and a Hall conductivity two times that of graphene. They will also occur in rombohedral multilayer graphene where the effective low energy model of the n layer compound has a Lifschitz dispersion relation $\omega\sim k^n$. 
Our mechanism for Hall viscosity will directly apply in these 2D (or effectively 2D) compounds.

In these multilayer systems formed by a family of isolated metallic 2D lattice planes indexed by a primitive reciprocal lattice vector $G^0$, the 3D Hall conductivity is
\beq
\sigma^{ab}_H=\frac{1}{2\pi}\sigma^{2D}_H\epsilon^{abc}G^0_c.
\label{3DH}
\eeq
Any 2D system having  a quantum Hall effect (QHE) will exhibit, when stacked to form a 3D material, a 3DQHE if the inter--plane coupling is sufficiently weak, as it is the case of the so called Van der Waals (VdW) systems.
When the 2D constituents support elastic gauge fields, a 3D Hall viscosity will arise whose coefficient  is not related to a quantum anomaly. The obvious example in this class is graphite
whose 3D Hall conductivity  has been described in \cite{BHetal07}. In a homogeneous magnetic field perpendicular to the layers and when the Fermi level is in the
3D gap, the Hall conductance is found to be (in units of the quantum of conductance) $\sigma_{xy}=2\frac{(2n+1)}{c_{0}}$ where $c_{0}$ is the c-axis coupling constant of graphite estimated to be $c_{0}=6.74 \AA$. The corresponding 3D Hall viscosity will have an in-plane component 
\beq
\eta^H_{1112}=2\frac{(2n+1)\beta^2}{\pi a^2 c_{0}},
\eeq
where $a=2.456$\AA  is the in-plane lattice constant of graphene.
Nowadays the family of VdW materials constitutes an intense area of research so we cannot limit ourselves to consider graphite or transition metal dichalcogenides. Hexagonal boron nitride (hBN) is another VdW system where these effects are potentially observable and even bulk black phosphorus whih has recently been found to have Weyl nodes under pressure, \cite{FTY15} will also have the proposed elastic response.


%
%
%
\begin{figure}
\includegraphics[width=7cm]{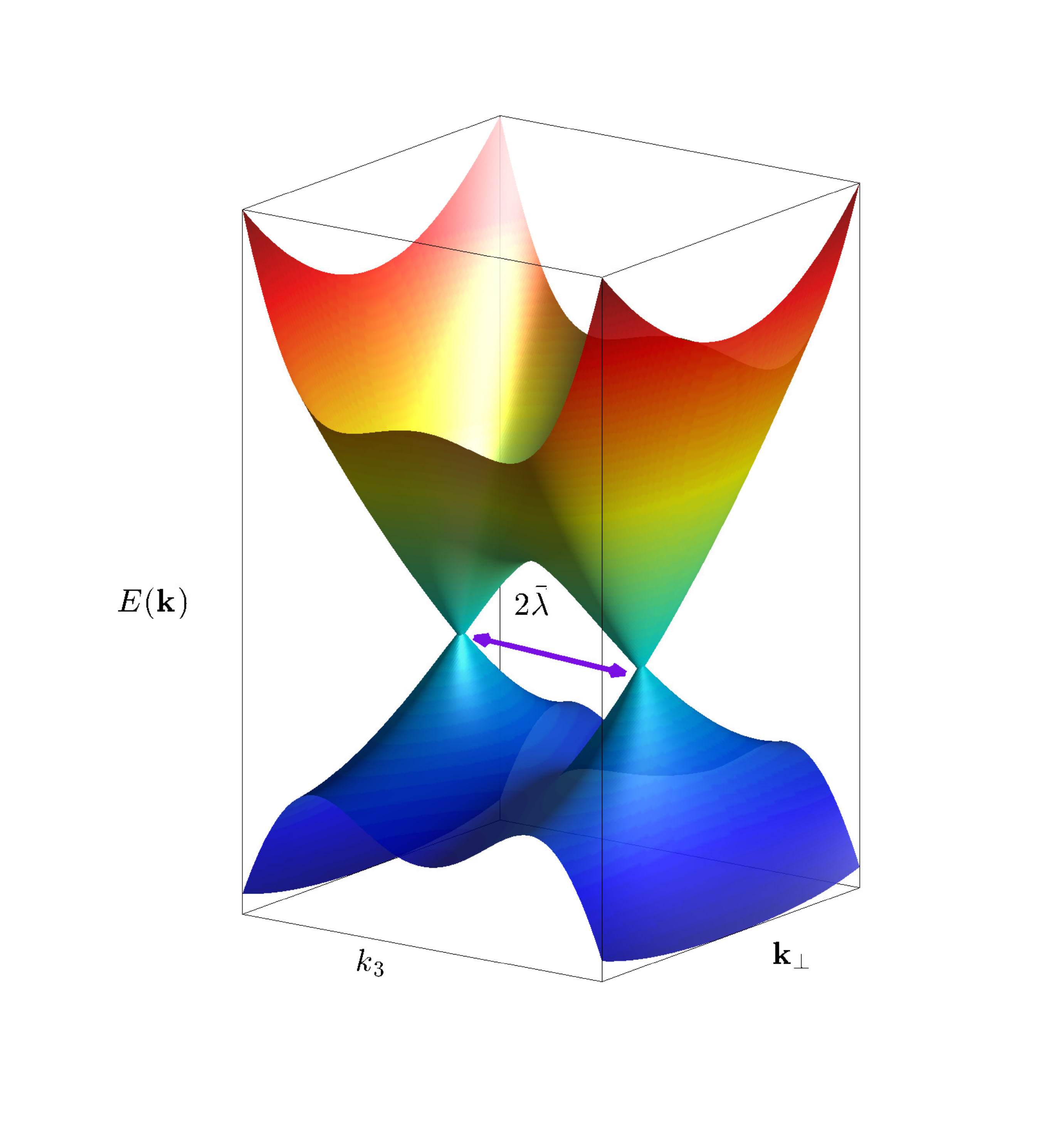}
\caption{\textbf{Band structure of representative Weyl semimetals}. Typical band structure of a Weyl semimetal with two nodal points at different points in the momentum space. The vector connecting the nodal points is a key ingredient in the axial anomaly.}
\label{WSMbands}
\end{figure}  
\subsection{Anomalous Hall viscosity in a Weyl semimetal}
The derivation of the Hall viscosity from Hall conductivity  can be extended to Weyl crystals in three spacial dimensions \cite{GSetal15}.  
For electronic states in a three-dimensional periodic
potential in a uniform magnetic field and when the Fermi level lies in an energy
gap, the Hall  conductivity adopts the form  \cite{Halp87}
$\sigma_{ij}=\frac{e^2}{2\pi c}\varepsilon_{ijk}G_k,$ 
where ${\bf G}$ is a vector in the reciprocal lattice of the periodic potential. The Hall current is given by ${\bf J}=\frac{e^2}{2\pi}{\bf E}\times{\bf G}$. The presence of elastic gauge fields will give rise in these 3D crystals,  to an anomaly related Hall viscosity coefficient $\eta_H$ obtained from the 3D Chern Simons term
\beq
S_{CS}=\int d^4 x G_i\epsilon^{ijkl}A^{el}_j\partial_k A^{el}_l.
\label{3DCS}
\eeq
For a general 3D crystal breaking time reversal symmetry, the anomalous hall effect is characterized
by a momentum space vector ${\vec \nu}$ called the Chern vector. The anomalous hall conductivity is given by the same expression:
\beq
\sigma_{ij} = \frac{e^2}{ 2\pi c} \epsilon_{ijk}\nu_k,
\label{3DHE}
\eeq
with the lattice vector replaced by ${\vec \nu}$. Our mechanism for the Hall viscosity will also apply in the three dimensional case by replacing the gauge fields in eq. \eqref{3DCS} by the  elastic gauge fields. 
The elastic Hall viscosity coefficient will be 
\beq
\eta_H\sim\frac{\beta^2}{ a^2 l },
\eeq
where $l$ is a characteristic length associated to the Chern vector. This is the basic mechanism at work. In (3+1) dimensions there are some subtleties concerning the coefficient of the Chern Simons action that will be discussed in detail in the Methods sec. \ref{elWSM}. In what follows we will concrete these general expressions working out the anomalous Hall viscosity of Weyl semimetals. 
 
Weyl semimetals (WSM) are the ``3D graphene": They break time reversal symmetry and have pairs of Weyl nodes of opposite chirality connected by Fermi arcs.  Their low energy description in terms of 3D Weyl fermions has given rise to an enormous amount of works related to quantum field theory physics: experimental realization of the axial anomaly, chiral magnetic effect, etc. (see \cite{HQ13,Bur15} and the references there in). To enhance the excitement,  WSM physics has been reported recently in several compounds, particularly  TaAs\cite{XuLiuetal15}. 
WSM whose Fermi surface crosses a given  set $i$ of Weyl nodes of chiralities $\xi_i$ located at positions ${\vec P}_i$  have an  anomalous Hall conductivity given by eq.  \eqref{3DHE} with a Chern vector  \cite{LY11} ${\vec \nu}=\Sigma_i (-1)^{\xi_i}{\vec P}_i$. Nevertheless the Hall viscosity of the intrinsic material at zero temperature and chemical potential is zero \cite{GMS14}.
By our proposed mechanism Weyl semimetals will have an anomalous Hall viscosity at zero temperature and zero chemical potential given by the coefficient dictated precisely by the chiral anomaly supplemented with the elastic constant of the material. 

A detailed derivation of the elastic gauge fields for a lattice model of a Weyl semimetal is given in   sec. Methods \ref{elWSM}. The low energy effective action in the continuum limit around the Weyl points ($\pm \bm{\lambda})$ is given by ($v=at$)
\begin{equation}
H_{W}(\bm{k})=\psi^{+}_{\pm,\bm{k}}\left(\bm{\sigma}(v\bm{k}_{\perp}\pm\bm{A}^{el}_\perp)\mp 
(v_{3}k_{3}\pm A_3^{el})\sigma_{3}\right)\psi_{\pm,\bm{k}}.
\end{equation}
%
The elastic gauge fields are proportional to the strain tensor $u_{ij}$ with coefficients that depend on
the parameters of the model:
\begin{eqnarray}
A^{el}_{1}=\beta \sqrt{b^{2}_{3}-m^{2}}\quad u_{31},\nonumber
\\
A^{el}_{2}=\beta\sqrt{b^{2}_{3}-m^{2}}\quad u_{32},\nonumber
\\
A^{el}_{3}=\beta\frac{r}{2}\frac{b^{2}_{3}-m^{2}}{t^{2}}u_{33}.
\end{eqnarray}

%
Integrating out the fermions in the four components formulation where the field ${\bm\lambda}$ couples as an axial field, we get through the electromagnetic chiral anomaly, the effective action
\begin{eqnarray}
&&\Gamma_{eff}[u]=\frac{1}{48\pi^{2}}\int d^4 x \epsilon^{\mu\nu\rho\sigma}\lambda_{\mu}A^{el}_{\nu}\partial_{\rho}A^{el}_{\sigma}=\nonumber\\
&=&\frac{\beta^{2}}{48\pi^{2}}\left(\frac{b^{2}_{3}-m^{2}}{v^{2}}\right)^{\frac{3}{2}}\int d^4 x
\left(u_{31}\dot{u}_{32}-u_{32}\dot{u}_{31}\right),
\label{CSWSM}
\end{eqnarray}
The numerical coefficient in (\ref{CSWSM}) is explained in sec. Methods \ref{elWSM}).
The anomalous Hall viscosity (in the absence of a magnetic field) is
\begin{equation}
\eta_{H}=\frac{\beta^{2}}{24\pi^{2}}\frac{1}{a^{3}}\Big(\frac{b^{2}_{3}-m^{2}}{t^{2}}\Big)^{\frac{3}{2}}.
\label{HVWSM}
\end{equation}
As it can be read from eq. \eqref{CSWSM} the  new Hall viscosity obtained from the elastic gauge coupling is a coefficient $\eta_{3132}$ that points along the direction of the vector $\bm \lambda$ breaking time reversal and rotation symmetries. Since in the perpendicular plane between the two nodal points the system is a Chern insulator (see Fig. \ref{WSMbands}), there will also be an in--plane component $\eta_{1112}$ coming from the electron-phonon mechanism described in ref. \cite{SHR15}.

\section{ Discussion and wrapping up}
%
%
%
%
%
%

In the present work, we have identified a new mechanism to generate a Hall viscosity and a large number of new materials and systems that will display such effect. As it can be understood from the previous general discussion\cite{BCQ12,SHR15}, it is natural to measure changes in the phonon structure due to the presence of $\eta^{H}$. In-plane phonon dispersion measurements can be performed by X-ray scattering\cite{MMD07}, Brillouin scattering\cite{WLN08}, and electron energy loss spectroscopy\cite{GD12}, all already done in the case of graphene and VdW materials. Also, since now we are also open to three dimensional electronic systems,  three dimensional probes like neutron scattering can in principle be also used. The Hall viscosity can be related to changes in the phonon dispersions in presence of an external magnetic field. 
In order to get some numbers we can take graphene as the paradigmatic example. Using standard values\cite{ZKF09} for the Lam\'{e} coefficients $\lambda=2$ eV\AA$^{-2}$ and $\mu=10$ eV\AA$^{-2}$ a characteristic frequency associated to the Hall viscosity can be defined, $\omega_{H}=\frac{\pi a^{2}\mu}{\beta^{2}\hbar}\simeq 95$ eV, which is a rather large frequency scale compared with the in-plane acoustic phonon frequencies in most systems ($\sim$ hundreds of meV). It intuitively means that the changes in the acoustic phonons will be hard to observe within current experimental resolutions at least in two dimensions. In spite of that, we believe that increasing the number of systems that display a Hall viscosity through elastic gauge fields, it is a matter of time to find an experimental probe that could detect and measure $\eta^{H}$.

We end up summarizing the main findings of this work and their physical implications:
\begin{itemize}
\item The elastic gauge fields encoding the interaction of the electronic properties o with lattice deformations of crystals 
are a very general phenomenon.  First described in graphene, they have also been obtained in the low energy effective models of other 2D systems as MoS$_2$ but there were no examples in 3D.  We have deduced the presence and structure of the elastic gauge fields in a 3D model of Weyl semimetals (Sec. Methods).

\item In any compound supporting elastic gauge fields Hall conductivity implies Hall viscosity. This is particularly important in 3D where the precise form of the gravitational and chiral anomalies excludes the possibility of standard Hall viscosity in the absence of torsion. For these systems with finite density or temperature where a Hall viscosity exists, the contribution to the Hall viscosity  from elastic gauge fields is several orders of magnitude bigger than that coming from phonons or metric deformations. 

\end{itemize}

The lattice aspects of the viscoelastic response of topological crystals have been recently analyzed in  \cite{SHR15}. Our new contribution to the Hall viscosity, although  not explicitly discussed, could certainly be worked out   as a part of their general analysis. What we emphasize here is the analogy of the elastic gauge fields with the standard electromagnetic partners which makes the connection with the anomaly related topological aspects  very transparent.

\acknowledgments 
We thank Carlos Hoyos for enlightening discussions on the Hall viscosity and Juan Ma\~nes for comments on the elastic gauge fields.  Special thanks go also to Jos\'e Silva-Guill\'en for help with the figures. This research was supported in part by the Spanish MECD grants FIS2011-23713, PIB2010BZ-00512, the  European Union structural funds and the Comunidad de Madrid MAD2D-CM Program (S2013/MIT-3007) 
and by the European Union Seventh Framework Programme under grant agreement no. 604391 Graphene Flagship.


\newcommand{\npb}{Nucl. Phys. B}\newcommand{\adv}{Adv.
  Phys.}\newcommand{\epl}{Europhys. Lett.}

\newpage
\section{Methods}
\subsection{Elastic gauge fields in graphene}
\label{elgraphene}

The fact that the lattice deformations couple to the low energy electronic excitations of graphene in the form of elastic gauge fields has been explained in a number of articles and reviews. Nevertheless, being central to our work, we will repeat the derivation here taken mostly from ref. \cite{VKG10}.
Elastic deformations can be introduced in a tight binding calculation simply changing the hopping parameters. In the case of non equal hopping parameters $\gamma_i=t_0+t_i$, one gets,  near the a
$K$ point, the effective low energy Hamiltonian: \cite{KM97,SA02b}
\begin{equation}
H=v_F\psi^{+}\sigma_{i}\left(i \partial_{i} -A^{el}_{i}\right) \psi,  
\label{df1}
\end{equation}
where 
$v_F=(3/2)ta$, $t$ is the hoping parameter of the undeformed lattice and $a$ is the lattice constant. The vector field $\vec{A}$ is
\begin{eqnarray}
A^{el}_1 &=&\frac{\sqrt{3}}{2v_F}\left(t_3-t_2\right) , \nonumber
\\
A^{el}_2 &=&\frac{1}{2v_F}\left( t_2+t_3-2t_1\right), \label{df2}
\end{eqnarray}
In the weakly deformed lattice, assuming that the atomic displacements $%
\vec{u}$ are small in comparison with the lattice
constant $a$ one has
\begin{equation}
t_i=t+\frac{\beta t}{a^2}\vec{\delta}%
_i\cdot\left( \vec{u}_i-\vec{u}_0\right) ,  \label{df3}
\end{equation}
where  $\vec{\delta }_i$ are the nearest-neighbor vectors,  $%
\vec{u}_0$ is the displacement vector for the central atom, and
\begin{equation}
\beta =-\frac{\partial \ln t}{\partial \ln a}\simeq 2 \label{df4}
\end{equation}
is the electron Gr\"{u}neisen parameter. The continuum limit (elasticity theory)  with a
displacement field $\vec{u}\left( \vec{r}\right) $ is performed by making the substitution
\begin{equation}
\vec{u}_i-\vec{u}_0\propto \left( \vec{\delta }_i\cdot\nabla \right)
\vec{u}\left( \vec{r}\right),
\end{equation}
from where we obtain the effective gauge field  \cite{SA02b,Ma07}:
\begin{eqnarray}
A^{el}_1 &=&\frac{\beta}{a}\left( u_{11}-u_{22}\right) ,  \nonumber
\\
A^{el}_2 &=&-2\frac{\beta}{a} u_{12}.  \label{df5}
\end{eqnarray}

For the other valley, $K^{\prime }$, the sign of the vector
potential (\ref {df5}) is opposite, in agreement with the requirement
of time-reversal invariance. 
 
The gauge field (\ref{df5}) is proportional to the
deformation tensor which is directly involved in the density of
elastic energy. This means that, although the coupling to the
fermions is gauge invariant,  the problem
as a whole is not. A pure gauge rotation 
by the gradient of a scalar function will change the kinetic term of the
elastic part.



\subsection{Emergent elastic vector fields in a Weyl semimetal.}
\label{elWSM}
To illustrate how emergent vector fields associated to elasticity appear in a Weyl semimetal phase we can consider the following simple model \cite{VF13,SHR15} of s-, and p-like electrons hopping in a cubic lattice and chirally coupled to an on-site constant vector field $\bm{b}$. The parameters t, r, and m, represent, in a tight-binding description, the hopping matrix elements between s and p states, hopping between the same kind of states, and the difference of on-site energies between s and p states, respectively. The vector field $\bm{b}$ breaks time reversal symmetry. Without loss of generality, we will choose the vector field $\bm{b}$ to point along the OZ direction.
The tight binding  Hamiltonian is:
\begin{equation}
H_{0}=\sum_{i,j}c^{+}_{i}\left(i t\alpha_{j}-r\beta\right)c_{i+ j}+(m+3r)\sum_{i}c^{+}_{i}\beta c_{i}+h.c,\label{Hzero}
\end{equation}
where $i$ labels the position $\mathbf{R}_{i}$ and $j$ labels the six next nearest neighbors $\mathbf{a}_{j}$ of length $a$ in the cubic lattice. The matrices $\alpha_{i}$ and $\beta$ are the standard Dirac matrices. In the unstrained situation we will set all the hopping terms $t$ equal for simplicity.

We will focus on the parameter regime $0>m>-2r$ corresponding to a topological insulating phase. The long wavelength limit of this model around the $\Gamma$ point ($\bm{k}=0$) is an isotropic massive Dirac system. Since $H_{0}$ describes a massive Dirac model in the continuum limit in three spatial dimensions, in order to find a non vanishing Hall viscosity we need to break both time reversal symmetry and the effective rotational symmetry in the continuum. In the lattice model we can do both operations by including an on-site axial coupling between the fermion fields and a constant vector field $\bm{b}$. Without loss of generality we will choose $\bm{b}$ to point along the third axis, with $b_{3}>m$:
\begin{equation}
H_{b}=\sum_{i} b_{3} c^{+}_{i}\alpha_{3}\gamma_{5}c_{i}.\label{Hb}
\end{equation}
\begin{figure*}
\includegraphics[width=\textwidth]{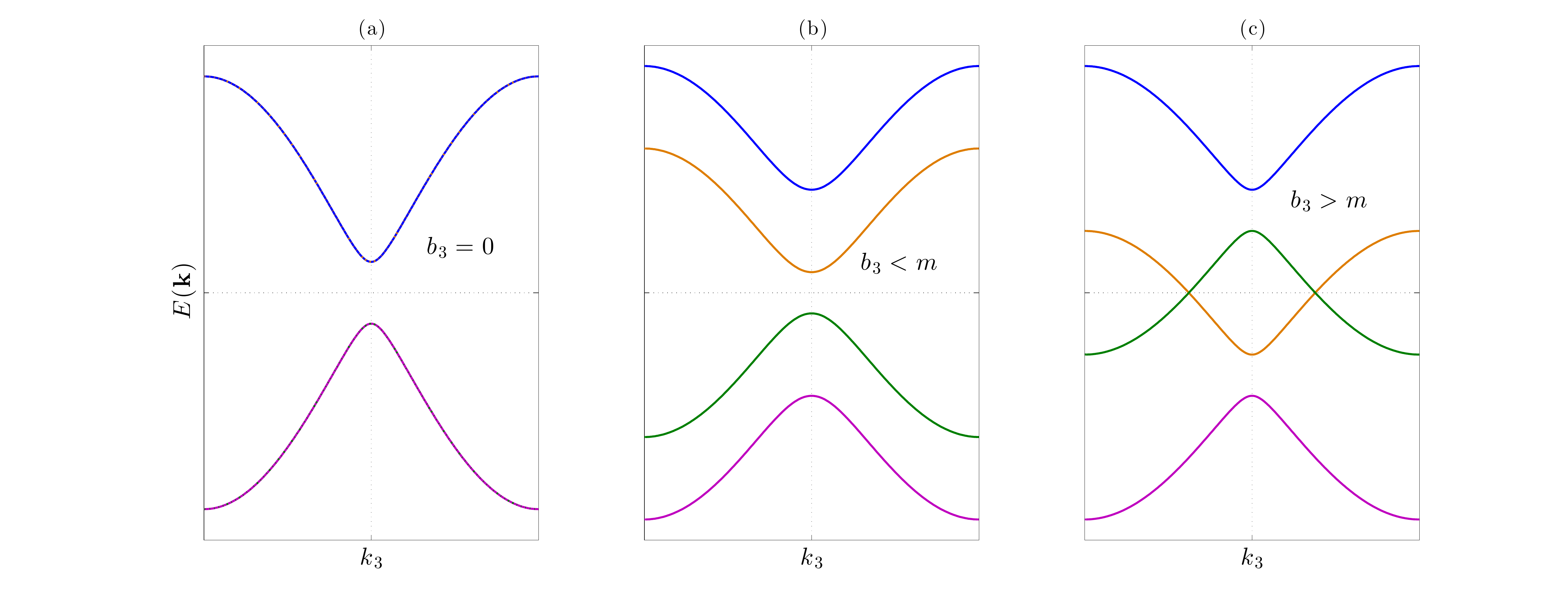}
\caption{\textbf{Evolution from a topologogical insulator to a Weyl semimetal}. For $b_{3}=0$ the spectrum consists in two pairs of degenerate bands due to time reversal symmetry, (a). When $0<b_{3}<m$ the band degeneracy breaks down and a high energy sector differentiates from a low energy sector, but the system is still gapfull, (b). When $b_{3}>m$, the low energy bands cross each other at two definite points in the Brillouin zone. At sufficiently low energies, the system consists in two pairs of Weyl fermions with opposite chirality, (c).}
\label{Hbbands}
\end{figure*}
With the choice $b_{3}>m$ the spectrum  of the total Hamiltonian (\ref{Hzero})$+$(\ref{Hb}) consists in two bands crossing at two Fermi points in the Brillouin Zone and two bands at higher and lower energies, as it can be seen in Fig.(\ref{Hbbands}). To find the low energy effective model around these two Fermi points, we shall proceed in two steps: we will expand the lattice model for small momenta compared with $1/a$, so after Fourier transforming (\ref{Hzero})$+$(\ref{Hb}) and introducing the following approximations $\sin(k_{j}a)\simeq k_{j}a$ and $\cos(k_{j}a)\simeq 1-\frac{k^{2}_{j}a^{2}}{2}$, one gets ($v=at$)
\begin{eqnarray}
H_{0}(\bm{k})&=&\sum_{j}c^{+}_{\bm{k}}\left(v\alpha_{j}k_{j}+m\beta+b_{3}\alpha_{3}\gamma_{5}\right)c_{\bm{k}}\equiv\nonumber\\
&\equiv& \sum_{j}c^{+}_{\bm{k}}\mathcal{H}_{0}(\bm{k})c_{\bm{k}}.
\end{eqnarray}
And now we project out the high energy bands. After a unitary transformation, the Hamiltonian matrix $\mathcal{H}_{0}(\bm{k})$ can be written as
\begin{eqnarray}
\mathcal{H}_{0}(\bm{k})=\left(\begin{array}{cc}
v\bm{\sigma}\cdot\bm{k}_{\perp}+(b_{3}+m)\sigma_{3} & vk_{3}\sigma_{3}\\
vk_{3}\sigma_{3} & v\bm{\sigma}\cdot\bm{k}_{\perp}+(b_{3}-m)\sigma_{3}\end{array}\right),
\end{eqnarray} 
with $\bm{k}_{\perp}=(k_{1},k_{2})$. The four-component wavefunction can be written in two-component blocks $\left(\phi_{\bm{k}},\psi_{\bm{k}}\right)$. For energies $E\ll m+b_{3}$ and momenta $k_{j}\ll (m+b_{3})/v$ we can write $\phi_{\bm{k}}\simeq-\frac{k_{3}}{m+b_{3}}\psi_{\bm{k}}$ and the effective two-band model takes the form:
\begin{equation}
H_{eff}(\bm{k})=\psi^{+}_{\bm{k}}\left(v\bm{\sigma}\cdot\bm{k}_{\perp}+\frac{1}{m+b_{3}}(b^{2}_{3}-m^{2}-v^{2}k^{2}_{3})\sigma_{3}\right)\psi_{\bm{k}}.\label{Heff}
\end{equation} 
From this effective Hamiltonian it is easy to see that the momenta where the two bands cross are $\bm{\lambda}_{\pm}=(0,0,\pm \sqrt{\frac{b^{2}_{3}-m^{2}}{v^{2}}})$. Expanding now around these two points $\bm{k}\simeq \bm{\lambda}_{\pm}+\delta\bm{k}$, the final Hamiltonian takes the form of two massless three dimensional Dirac fermions $\psi_{\pm}$ separated $2\bm{\lambda}$ in the momentum space:
\begin{equation}
H_{W}(\delta\bm{k})=\psi^{+}_{\pm,\bm{k}}\left(v\bm{\sigma}\cdot\delta\bm{k}_{\perp}\mp v_{3}\delta k_{3}\sigma_{3}\right)\psi_{\pm,\bm{k}},
\end{equation}
with $v_{3}=2v\sqrt{\frac{b_3-m}{b_3+m}}$.

Now let us focus on the part of the original lattice Hamiltonian that depends on the strain. As discussed in the case of graphene,
the strain tensor $u_{ij}$ enters in the tight binding approach through the change of the hopping parameters $t$ when the lattice is distorted. The general recipe is (see also  ref. \cite{SHR15}) for $t_{1}$, 
\beq
t\alpha_{1}\rightarrow t(1-\beta u_{11})\alpha_{1}+t \beta u_{12}\alpha_{2}+t\beta u_{13}\alpha_{3}, 
\eeq
and the same for $t_{1}$ and $t_{2}$, and 
\begin{equation}
r\rightarrow r_{j}\simeq r(1-\beta u_{jj}),
\end{equation}
being $\beta$ the corresponding Gr\"{u}neisen parameter of the model. Inserting these hopping changes in the original Hamiltonian (\ref{Hzero}), we can define the strained Hamiltonian as the sum of the original Hamiltonian $H_{0}$ and the strain dependent part $H[u_{ij}]$. For our purposes we will first focus on these two parts of $H[u_{ij}]$:
\begin{equation}
\delta H_{1}[u_{ij}]=t\beta \sum_{\bm{k}}c^{+}_{\bm{k}}[\left(u_{31}\alpha_{1}+u_{32}\alpha_{2}-u_{33}\alpha_{3}\right)\sin(k_{3}a) ]c_{\bm{k}},
\end{equation}
and
\begin{equation}
\delta H_{2}[u_{ij}]=r\beta\sum_{\bm{k}}u_{33}(1-\cos(k_{3}a))c^{+}_{\bm{k}}\beta c_{\bm{k}}.
\end{equation}
Projecting out the high energy sector and expanding around the two nodal points $\lambda_{\pm}$, the strain dependent Hamiltonian part takes the form
\begin{eqnarray}
\delta H[u_{ij}]&=&\pm \lambda\beta v \sum_{\bm{k}}\sum_{j=1,2}u_{3j}\psi^{+}_{\pm,\bm{k}}\sigma_{j}\psi_{\pm,\bm{k}}-\nonumber\\
&-&\frac{r a^{2}\beta}{2}\lambda^{2}u_{33}\psi^{+}_{\pm,\bm{k}}\sigma_{3}\psi_{\pm,\bm{k}}.
\end{eqnarray}
We have found that, around the two nodal points, strain couples to the low energy electronic sector as a vector field:

\begin{eqnarray}
A^{el}_{1}=\beta \sqrt{b^{2}_{3}-m^{2}} u_{31},\nonumber
\\
A^{el}_{2}=\beta\sqrt{b^{2}_{3}-m^{2}} u_{32},\nonumber
\\
A^{el}_{3}=\beta\frac{r}{2}\frac{b^{2}_{3}-m^{2}}{t^{2}}u_{33}.
\label{3Delasticfield}
\end{eqnarray}
We also note that this coupling is chiral, that is, the elastic vector field $\bm{A}^{el}$  couples with opposite signs to the electronic excitations around the two Weyl nodes. What we have found is that, when the cubic symmetry in the original lattice model, or the continuous rotation symmetry in the effective model is broken by a constant vector $\bm{b}$, strain couples to electrons as a chiral vector field, similarly to what happens in graphene or other two dimensional systems.

Integrating out fermions to get an effective theory for the elastic distortions will give rise, through the electromagnetic chiral anomaly, to a term of the form
\begin{eqnarray}
\Gamma[u_{ij}]&=&\frac{1}{3}\frac{\beta^{2}\lambda}{16\pi^{2}}\int d^3 x \left(A^{el}_{1}\dot{A}^{el}_{2}-A^{el}_{2}\dot{A}^{el}_{1}\right)=\nonumber\\
&=&\frac{1}{3}\frac{\beta^{2}\lambda^3}{16\pi^{2}}\int d^3 x \left(u_{31}\dot{u}_{32}-u_{32}\dot{u}_{31}\right),\label{HVaction}
\end{eqnarray}
and the Hall viscosity associated to the chiral anomaly mechanism is (in units $\hbar=1$)
\begin{equation}
\eta_{H}=\frac{\beta^{2}}{24\pi^{2}}\frac{1}{a^{3}}\Big(\frac{b^{2}_{3}-m^{2}}{t^{2}}\Big)^{\frac{3}{2}}.
\end{equation}

\subsection{Chiral anomaly and Hall viscosity in Weyl semimetals}
\label{3dAnomaly}
\begin{figure*}
(a)
\includegraphics[width=6cm]{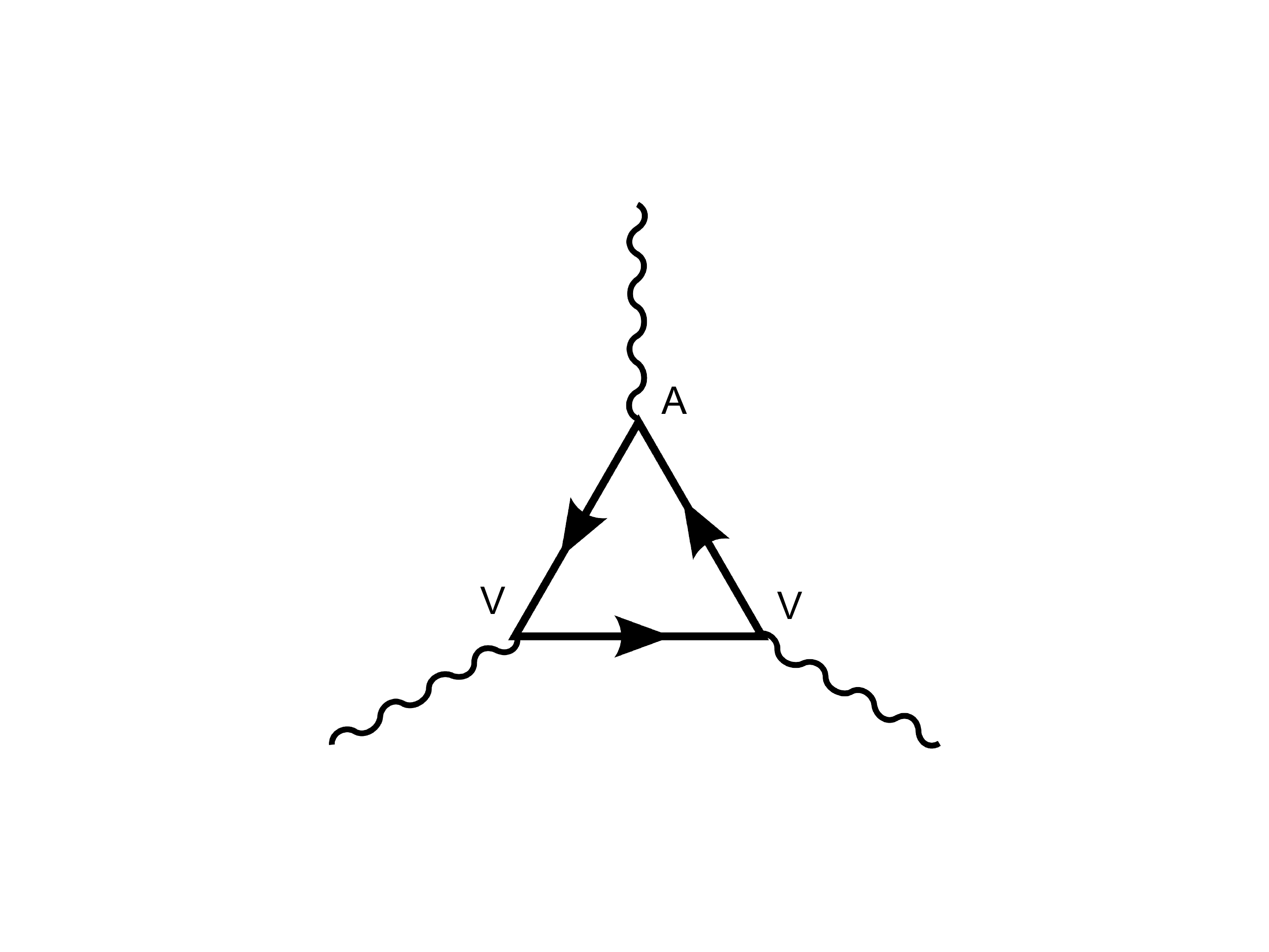}
(b)
\includegraphics[width=6cm]{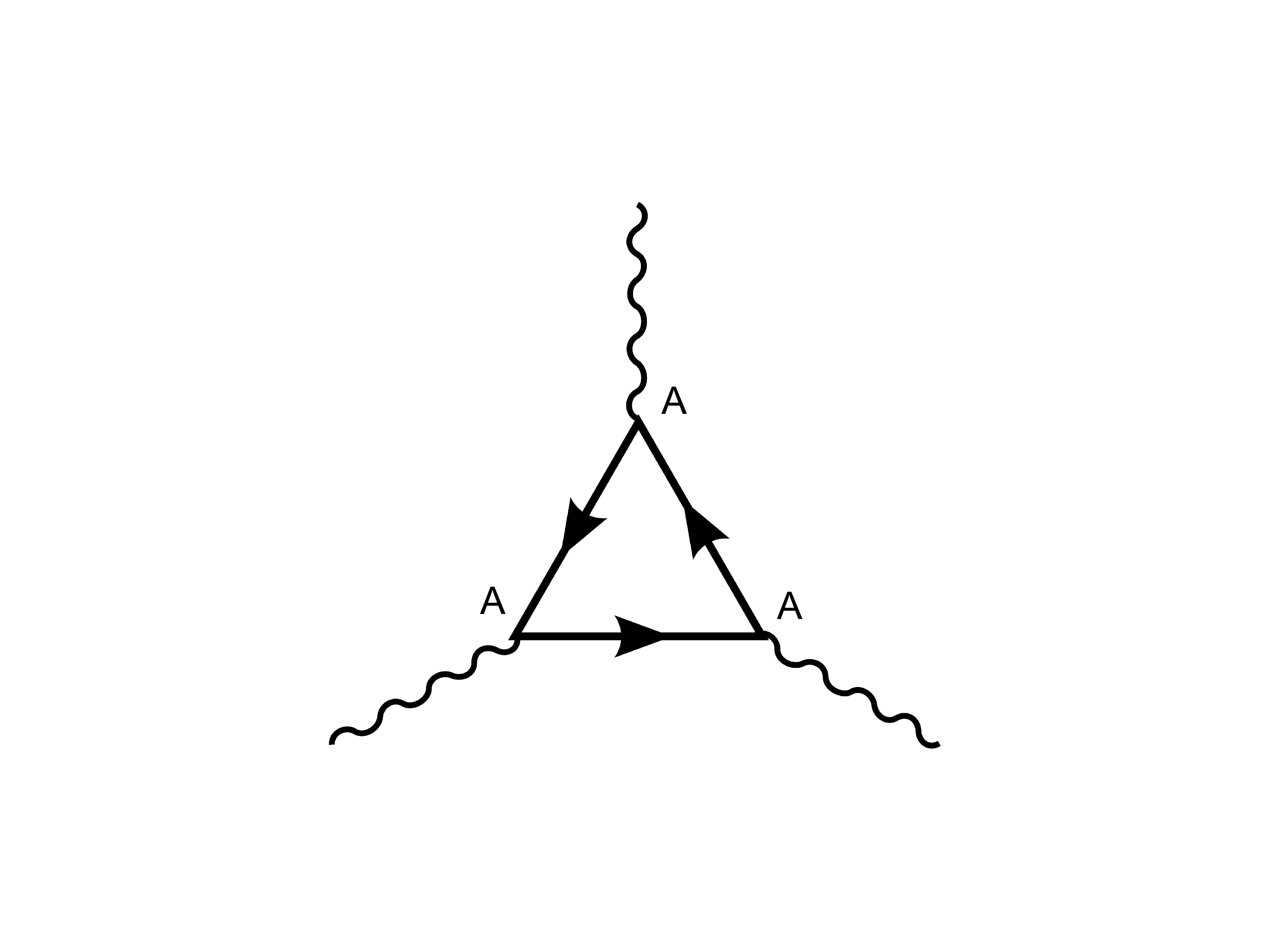}
\caption{\textbf{Anomalous Feynman diagrams.} When both axial and vector fields are coupled to Weyl fermions, there are two diagrams that contribute to the axial anomaly. Imposing that the electromagnetic current must be conserved forces these two amplitudes to not be equal being the proportionality factor $\frac{1}{3}$. }
\label{Fig:Anomaly}
\end{figure*}
Let us consider he action for a massless Dirac fermion  coupled to a U(1) vector $A_\mu$ and to a constant axial vector $b_{\mu}$  in (3+1) dimensions:
\beq
S=\int d^4 x \bar\psi\gamma^{\mu}(i\partial_\mu+e A_\mu+b_{\mu}\gamma^{5})\psi.\label{Fullaction}
\eeq
At the classical level the action is invariant separately under vector and axial transformations: 
$\psi\to \exp[i\alpha]\psi$,   $\psi\to \exp[i\theta\gamma_5]\psi$ what ensures the separate conservation of the  vector $J_\mu=\bar\psi\gamma_\mu\psi$ and axial $J_\mu^5=\bar\psi\gamma_\mu\gamma_5\psi$ currents: $\partial^\mu J_\mu=0$, $\partial^\mu J_\mu^5=0$. Also, since we have chosen $b_{\mu}$ to be constant, classically we can make a change in the chiral phase of the fermion fields to eliminate it from the action (\ref{Fullaction}). As we know well, we cannot remove completely the vector $b_{\mu}$ from our theory at the quantum level due to the chiral anomaly\cite{Grushin12,GT13}, that translates into the presence in the effective action for the electromagnetic field of the following term:
\begin{equation}
\Gamma_{a}[A]=\frac{e^2}{16\pi^{2}}\int d^{4}x \left(b_{\sigma}x^{\sigma}\right)\epsilon^{\mu\nu\alpha\beta}F_{\mu\nu}F_{\alpha\beta}\label{AAnomaly}.
\end{equation}
Integrating by parts and neglecting boundary terms, one can recognize in (\ref{AAnomaly}) the so-called Bardeen counter-term\cite{B69}. From the action (\ref{AAnomaly}), one easily obtains the anomalous current conservation laws at the quantum level: $\partial^{\mu}J_{\mu}=0$ and $\partial^{\mu}J^{5}_{\mu}=\frac{1}{16\pi^{2}}\epsilon^{\mu\nu\alpha\beta}F_{\mu\nu}F_{\alpha\beta}$.

Things become more involved when, besides the field $A_{\mu}(x)$ and the vector $b_{\mu}$, we consider adding to (\ref{Fullaction}) an axial vector field $A^{5}_{\mu}$ that couples to the chiral current $J^{5}_{\mu}$. In this case the chiral current is not only coupled to a constant vector that leads to ({\ref{AAnomaly}) after a chiral gauge transformation. In this case, the new effective action can be obtained by computing the amplitudes diagrammatically depicted in Fig.(\ref{Fig:Anomaly})\cite{BMN82,HZR14}. Up to quadratic terms in $A_{\mu}$ and $A^{5}_{\mu}$, the effective action now has an extra term: 
\beq
\Gamma_{a}[A^{5}]=\frac{1}{3}\frac{1}{16\pi^{2}}\int d^{4}x (b_{\sigma}x^{\sigma})\epsilon^{\mu\nu\alpha\beta}f_{\mu\nu}(x)f_{\alpha\beta}(x),\label{AAnomaly2}
\eeq
where 
$f_{\mu\nu}$ is the field strength associated to $A^{5}_{\mu}$. The relative factor of $1/3$ in the coefficients of the terms in (\ref{AAnomaly2}) and \eqref{AAnomaly}  is fixed by the requirement of gauge invariance $\partial^{\mu}J_{\mu}=0$. 

We are now ready to obtain the term in the effective action for elasticity corresponding to the Hall viscosity from (\ref{AAnomaly2}) by identifying the vector field $b_{\mu}$ with $(0,\bm \lambda)$ and $A^{5}_{\mu}$ as the elastic vector field in (\ref{3Delasticfield}). According to our discussion on the elastic gauge fields, since they are axial vector fields, and so is the vector $\vec b$, the coefficient of the effective action corresponds to the AAA vertex. This is the value that gives the final result for the anomalous Hall viscosity of the Weyl semimetal eq.(\ref{HVWSM}). 
\end{document}